\documentclass[submission, Proceedings]{SciPost}

\hypersetup{
    colorlinks,
    linkcolor={red!50!black},
    citecolor={blue!50!black},
    urlcolor={blue!80!black}
}

\usepackage[bitstream-charter]{mathdesign}
\DeclareSymbolFont{usualmathcal}{OMS}{cmsy}{m}{n}
\DeclareSymbolFontAlphabet{\mathcal}{usualmathcal}

\begin{document}

\begin{center}{\Large \textbf{
Transverse Single-Spin Asymmetries of Midrapidity Direct Photons and Neutral Mesons at PHENIX\\
}}\end{center}

\begin{center}
Nicole Lewis for the PHENIX Collaboration\textsuperscript{1}
\end{center}

\begin{center}
{\bf 1} University of Michigan
\\

* nialewis@umich.edu
\end{center}

\begin{center}
\today
\end{center}

% For convenience during refereeing (optional),
% you can turn on line numbers by uncommenting the next line:
%\linenumbers

\definecolor{palegray}{gray}{0.95}
\begin{center}
\colorbox{palegray}{
  \begin{tabular}{rr}
  \begin{minipage}{0.1\textwidth}
    \includegraphics[width=22mm]{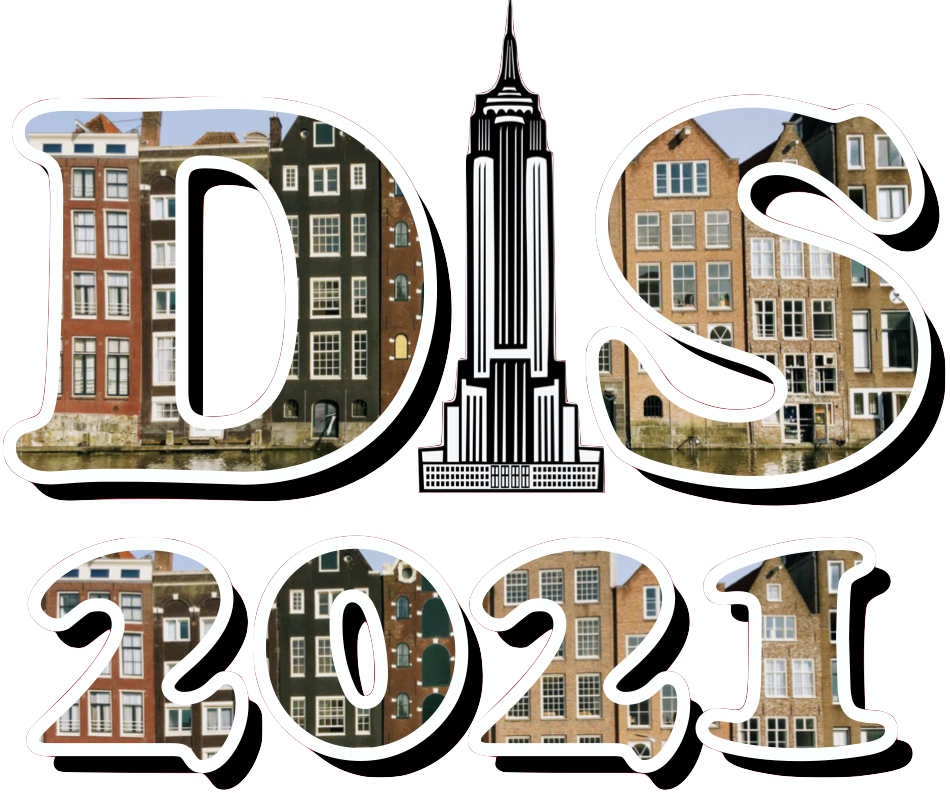}
  \end{minipage}
  &
  \begin{minipage}{0.75\textwidth}
    \begin{center}
    {\it Proceedings for the XXVIII International Workshop\\ on Deep-Inelastic Scattering and
Related Subjects,}\\
    {\it Stony Brook University, New York, USA, 12-16 April 2021} \\
    \doi{10.21468/SciPostPhysProc.?}\\
    \end{center}
  \end{minipage}
\end{tabular}
}
\end{center}

\section*{Abstract}
{\bf
%The abstract is in boldface and should fit in 8 lines.
%It should be written in a clear and accessible style, emphasizing the context, the problem(s) studied, the methods used, the results obtained, the conclusions reached, and the outlook. You can add a table contents, recommended if your paper is more than 6 pages long.
Results are presented for the transverse single-spin asymmetries of direct photons, neutral pions, and eta mesons for $|\eta|<0.35$ from $p^\uparrow + p$ collisions with $\sqrt{s} = 200$ GeV at PHENIX. As hadrons, $\pi^0$ and $\eta$ mesons are sensitive to both initial- and final-state effects and at midrapidity probe the dynamics of gluons along with a mix of quark flavors. Because direct photon production does not include hadronization, the direct photon TSSA is only sensitive to initial-state effects and at midrapidity provides a clean probe of the gluon dynamics in transversely polarized protons. All three of these results will help constrain the collinear twist-3 trigluon correlation function as well as the gluon Sivers function, improving our knowledge of spin-dependent gluon dynamics in QCD.
}

% TO DO: include a table of contents (optional)
% Guideline: if your paper is longer than 6 pages, include a TOC
\vspace{10pt}
\noindent\rule{\textwidth}{1pt}
\tableofcontents\thispagestyle{fancy}
\noindent\rule{\textwidth}{1pt}
\vspace{10pt}

\section{Introduction}
\label{sec:intro}
% TO DO: write your article here.
Transverse single-spin asymmetries (TSSAs) are a type of spin-momentum correlation measurement in hadronic collisions.  In the context of proton-proton collisions, one of the protons is transversely polarized while the other is unpolarized. The TSSA measures the asymmetry of particle yields which travel to the left versus the right of the polarized-proton-going direction.  Theoretical calculations which only include effects from high energy partonic scattering predict that spin-momentum correlations like these should be small, on the order of less than one percent~\cite{GordyPaper}, but in fact large nonzero asymmetries have been measured in a variety of collision systems.  This includes the forward $\pi^0$ asymmetry for $p+p$ and $p+A$ collisions with $\sqrt{s_{NN}} = 200~\rm{GeV}$ and transverse momentum up to $p_T \approx 7~\rm{GeV}/c$~\cite{STARforwardPi0}.  
%up to 40% at forward rapidity?
Because the perturbative part of QCD calculations could not account for these large spin-momentum correlations, this led to the reexamination of  the nonperturbative part.  Two theoretical frameworks were developed: transverse momentum dependent functions and collinear twist-3 correlation functions both of which describe spin-momentum correlations within the nucleon and in the process of hadronization.  

Traditionally parton distribution functions (PDFs) and fragmentation functions (FFs) are collinear, meaning that they integrate over the nonperturbative dynamics of partons and only depend on longitudinal momentum fractions.  Transverse momentum dependent (TMD) functions, as the name suggests, depends explicitly on the parton's relative transverse momentum $k_{T}$.  The Sivers function is a TMD PDF that describes the spin-momentum correlation between the transversely polarized proton and the nonperturbative transverse momentum of the quark~\cite{SiversFunction}.  The quark Sivers functions have been extracted through polarized semi-inclusive deep inelastic scattering (SIDIS) measurements, while the gluon Sivers function has remained comparatively unconstrained because SIDIS is not sensitive gluons at leading order~\cite{gluonsSiversFromSIDIS}.  The Collins function is an example of a TMD FF and describes the spin-momentum correlation between the transverse spin of a quark and the soft-scale relative transverse momentum of the unpolarized hadron it produces~\cite{CollinsFunction}.  In order for TMD factorization to apply, this transverse momentum must be both nonperturbative and much smaller than the hard-scale energy of the scattering event.  Thus, the most straight forward way to extract these TMD functions is with a two-momentum scale scattering process, like SIDIS.   In SIDIS it is also possible to isolate the effects from particular TMD functions through angular moments and measure both the Sivers~\cite{SiversInSIDIS1, SiversInSIDIS2} and Collins~\cite{CollinsInSIDIS} functions directly.   

However, the measurements that are presented in these proceedings are single scale processes and are measured as a function of transverse momentum which is used as a proxy for the hard scale.  In order to apply TMD functions to these measurements, one must take the $k_T$ moment of these functions.  The neutral meson TSSA presented in this document are sensitive to both initial- and final-state effects.  In the TMD picture these are divided into the Sivers effect where the $k_T$ moment of the Sivers function has been taken and the Collins effect where the Collins function has been convolved with the collinear Transversity function, a PDF that describes the correlation between the transverse spin of a quark and the transverse spin of the proton.  Theoretical calculations using TSSAs measured in proton-proton data have suggested that the Collins effect's contribution~\cite{CollinsContribution} is smaller than Sivers effect's contribution~\cite{SiversContribution}, which is consistent with the small Collins asymmetry that was measured for forward-rapidity $\pi^{0}$s in jets~\cite{pi0InJet}. Because the Sivers function is parity-time odd, in order to be nonzero, it must include a soft-gluon exchange with the proton fragment, which can happen before and/or after the hard-scattering event.  In processes like proton-proton to hadron interactions soft gluon exchanges are possible in both the initial- and final-state simultaneously, leading to the prediction of TMD factorization breaking~\cite{TMDFactorizationBreaking}.

Another approach towards describing TSSAs are collinear twist-3 correlation functions.  While traditional nonperturbative functions are twist-2 and only consider the interactions of a single parton in the proton and a single parton hadronizing at a time, twist-3  functions are multiparton correlation functions.  They describe the quantum mechanical interference between scattering off of one parton at a given $x$ versus scattering off of a parton of the same flavor and same $x$ plus an additional gluon.  These functions are broken into two types: the quark-gluon-quark ($qgq$) correlation functions describe the quantum mechanical interference between scattering of a single quark versus scattering off of a quark and a gluon, while the trigluon ($ggg$) correlation functions describe the interference between scattering off of one gluon versus scattering off of two.   Collinear twist-3 correlation functions are used to describe spin-momentum correlations both from initial-state proton structure and also from final-state hadronization.   Efremov-Teryaev-Qiu-Sterman (ETQS) function is a $qgq$ correlation function for the polarized proton~\cite{ ETQS1, ETQS2, ETQS3 }  that is related to the $k_T$ moment of the Sivers TMD PDF~\cite{TMDandTwist3Related}. Collinear $qgq$ correlation functions have been used to describe forward $\pi^0$ TSSA measurements by including contributions from both the ETQS function and final state hadronization effects, which dominate~\cite{Twist3FinalState, TMDandTwist3SameOrigin}.
%simultaneous fit indicates all TSSA have a common origin ~\cite{TMDandTwist3SameOrigin}
Twist-3 collinear functions have the added benefit that they do not depend on a soft-scale momentum and are uniquely suited to describing TSSA in proton-proton collisions where only one final state particle is measured.

\section{Experimental Setup}
These measurements were taken at the Relativistic Heavy Ion Collider (RHIC), the only collider in the world that is able to collide polarized proton beams. Both beams are polarized, and the polarization direction changes direction bunch-to-bunch in order to avoid systematic effects.  The polarization is maintained by a series of spin-rotating helical dipoles called Siberian snakes which are placed at diametrically opposite points along both RHIC rings. They flip the polarization direction for each bunch by $180^o$ degrees without distorting the trajectory of the beam, causing additive depolarization effects to cancel out.

PHENIX is one of the large multipurpose detectors around the RHIC ring.  In 2015 PHENIX took a transversely polarized proton-proton data set with an integrated luminosity of $60~\rm{pb}^{-1}$.  These measurements use photons that were measured in the electromagnetic calorimeter (EMCal) which has a pseudorapidity acceptance of $|\eta| < 0.35$ and two nearly back-to-back arms that each cover $\Delta\phi = \pi/2$ in azimuth.  The EMCal is comprised of eight sectors, six of which are made of lead scintillator towers and two of which are made of lead glass.  Events with high $p_T$ photons are selected though an EMCal-based high energy photon trigger.  

Traditionally TSSAs are measured as function of $\phi$ and then fit to a sinusoid to extract the amplitude.  This becomes more difficult with the limited PHENIX acceptance, especially for midrapidity asymmetries that tend to be consistent with zero.  So PHENIX midrapidity TSSA analyses integrate over the full $\phi$ range of the detector:
\begin{equation}
A_N  = \frac{\sigma_L - \sigma_R}{\sigma_L + \sigma_R} = \frac{1}{P\langle \mid \cos{\phi} \mid \rangle }\frac{N^\uparrow - \mathcal{R} \, N^\downarrow}{N^\uparrow + \mathcal{R} \, N^\downarrow}
\end{equation}
\noindent The acceptance correction, $\langle \mid \cos{\phi} \mid \rangle$, is used to correct for the dilution of the asymmetry over the EMCal $\phi$ range.  $\langle \mid \cos{\phi} \mid \rangle$ is measured as a function of $p_T$ for the $\pi^0$ and $\eta$ analyses since the diphoton azimuthal acceptance changes with the photon decay angle.  The asymmetry is also diluted by the beam not being 100\% polarized and so the asymmetry is divided the beam polarization, $P$ which  for this data set was 59\% on average.  This asymmetry formula takes advantage of the fact that the beam polarization direction changes bunch-to-bunch, where $N^\uparrow$ and $N^\downarrow$ are the particle yields for when the beam is polarized up and down, respectively.  Because this formula takes the ratio of particle yields from the same detector, effects from detector acceptance and efficiencies cancel out.  This asymmetry needs to be corrected for the relative luminosity of the different beam configurations:  $\mathcal{R} = \mathcal{L}^\uparrow / \mathcal{L}^\downarrow$.  This is calculated by taking the ratio of the number of events that fired the minimum bias trigger when the beam was polarized up divided by the same for when the beam was polarized down.  

\begin{figure}[h]
\centering
\includegraphics[width=0.7\textwidth]{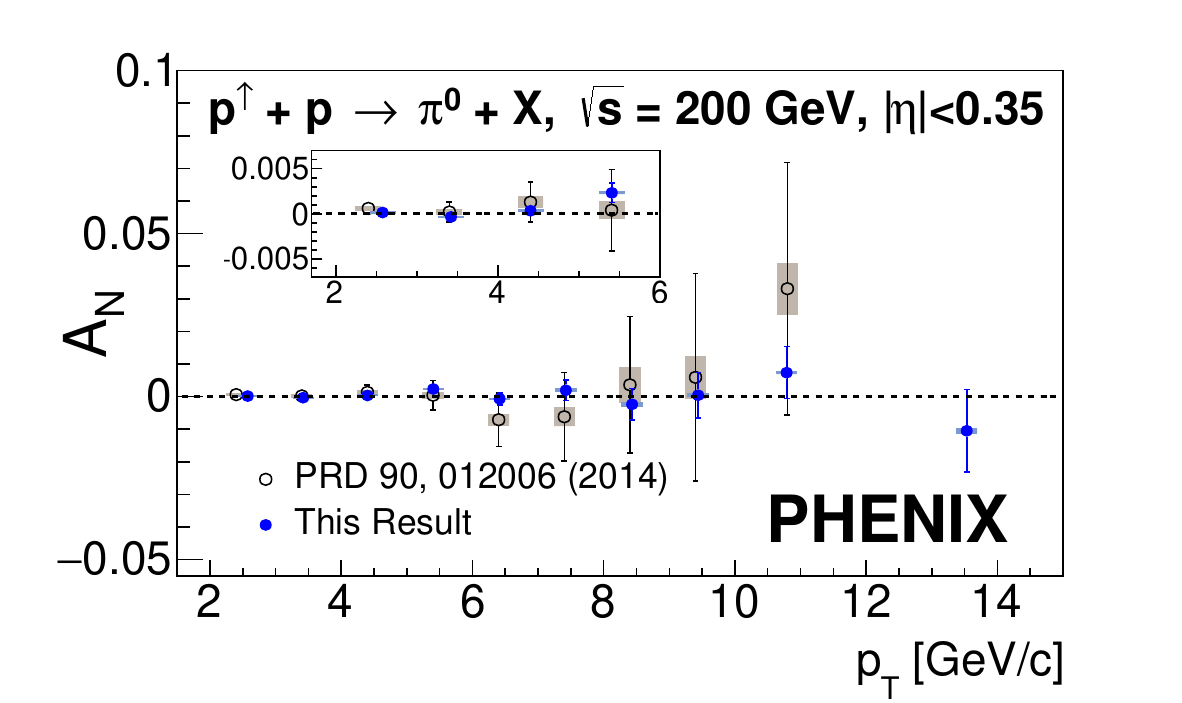}
\includegraphics[width=0.7\textwidth]{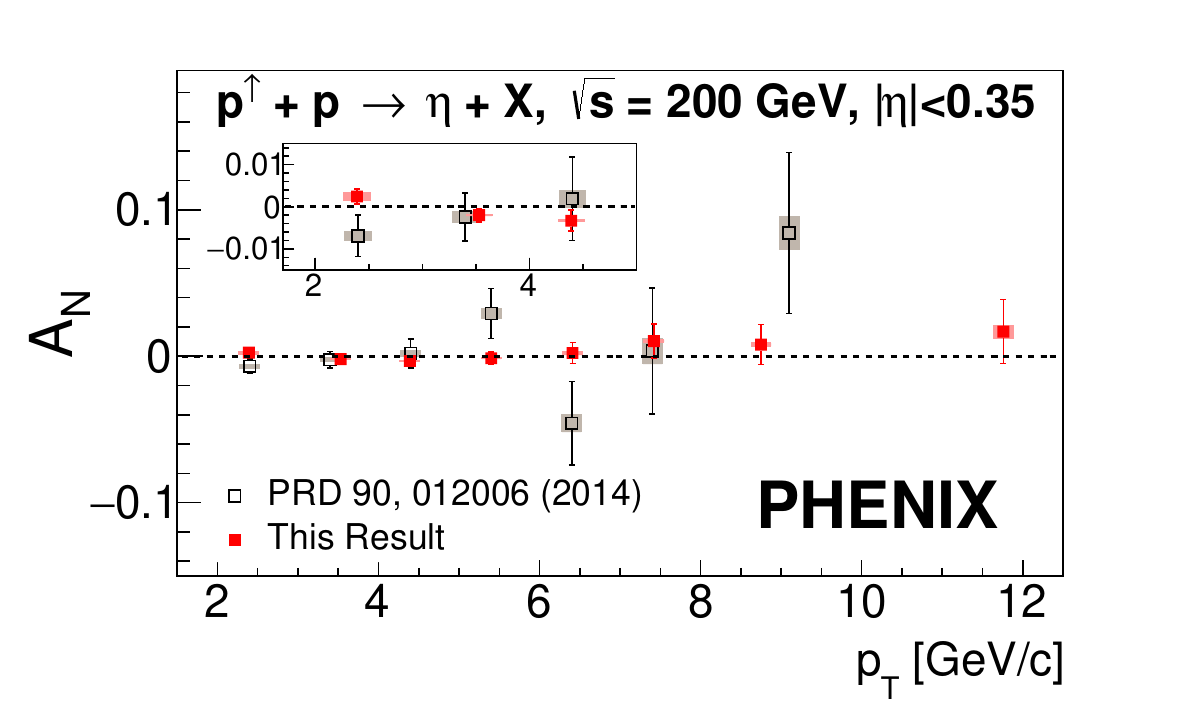}
\caption{The new $\pi^0$ and $\eta$ TSSA results~\cite{PPG234} plotted with the previously published PHENIX results~\cite{PPG135}.  An additional scale uncertainty of 3.4\% due to the uncertainty in the polarization is not shown}
\label{pi0AndEta}
\end{figure}

\section{$\pi^0$ and $\eta$ Meson TSSA}
Forward-rapidity light hadron production corresponds to probing the proton at higher $x$, meaning that valence quark spin-momentum correlations dominate the forward rapidity $\pi^0$ and $\eta$ TSSA measurements.  Measurements of the forward $\pi^0$ TSSA have been used to constrain $qgq$ correlation functions both in the transversely polarized proton and the process of hadronization~\cite{Twist3FinalState, TMDandTwist3SameOrigin}.  In contrast, at midrapidity the proton is probed at comparatively more moderate $x$, and so light hadron production includes contributions from valance quarks, gluons, and sea quarks~\cite{pi0AndEtaProduction}.  While this makes interpreting results a midrapidity more challenging, it does mean that these midrapidity TSSA measurements are sensitive to gluon dynamics at leading order.   As hadrons, $\pi^0$ and $\eta$ mesons are sensitive to both proton structure as well as in the process of hadronization and the $\eta$ meson is also sensitive strange quark dynamics.  The $\pi^0$ yields are comprised of photon pairs with invariant mass in the signal region $\pm 25~\rm{MeV}/c^2$ from the $\pi^0$ mass peak and $\eta$ meson yields are measured in the range $\pm 70~\rm{MeV}/c^2$ around the $\eta$ mass peak.  Figure~\ref{pi0AndEta} shows the new midrapidity  $\pi^0$ and $\eta$  TSSA results ~\cite{PPG234} compared with the previously published PHENIX results~\cite{PPG135}.  The $\pi^0$ asymmetry is consistent with zero to within $10^{-4}$ at low $p_T$ and the $\eta$ asymmetry is consistent with zero to within 0.5\% at low $p_T$.  Both new results have a statistical uncertainty that is three times smaller than the previous PHENIX results and have a higher reach in $p_T$.   
%comparison of pi0 and eta study the difference in isospin, strangeness, and mass

\begin{figure}[h]
\centering
\includegraphics[width=0.7\textwidth]{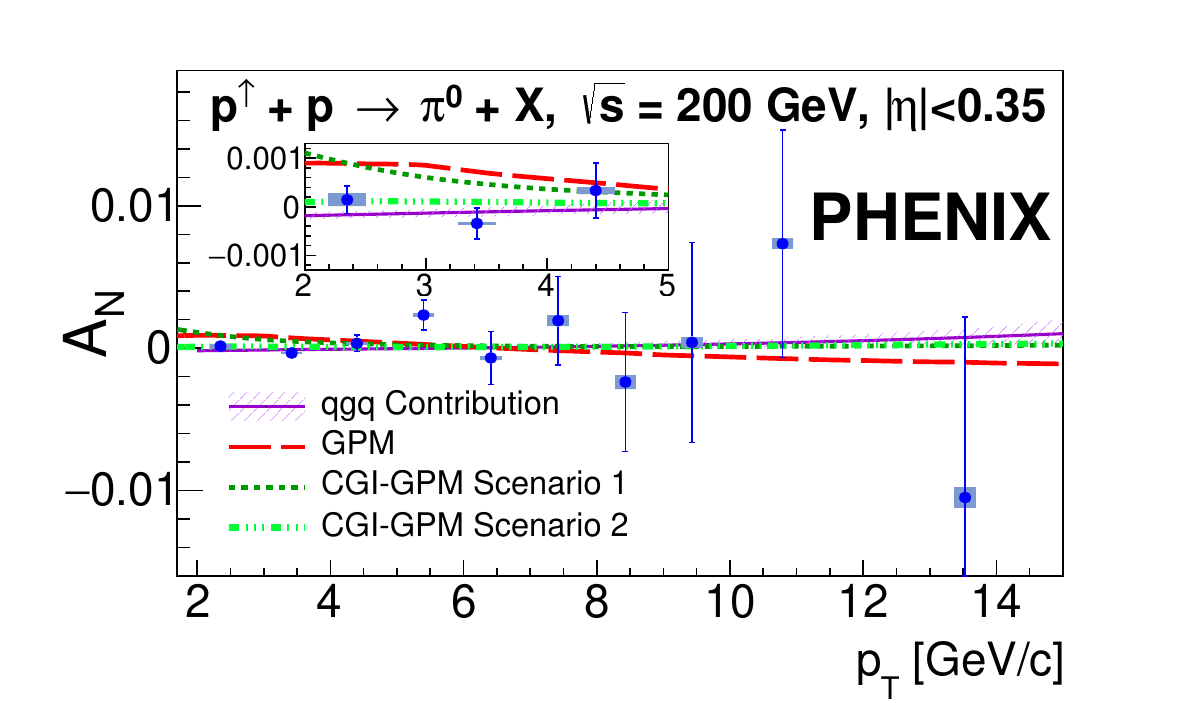}
\caption{The $\pi^0$ TSSA~\cite{PPG234} with theoretical curves in both the collinear twist-3~\cite{TMDandTwist3SameOrigin} and TMD~\cite{gluonSivers} frameworks.  See text for details.   }
\label{pi0Theory}
\end{figure}

Figure~\ref{pi0Theory} shows this  same $\pi^0$ result plotted with theoretical predictions.  The $qgq$ curves shows the predicted contribution of the quark-gluon-quark correlations from both the transversely polarized proton and the process of hadronization.  This calculation uses fits to data that were published in Ref~\cite{TMDandTwist3SameOrigin} and recalculated for the pseudorapidity ranges of this measurement.  Midrapidity $\pi^0$ production also includes a large fractional contribution from gluons, and so a full twist-3 collinear description of the midrapidity $\pi^0$ TSSA  also needs to include the contribution from the trigluon correlation function, such as those published in Ref~\cite{gggPi0}.  Given the small expected contributions from the $qgq$ correlation functions, this measurement can constrain future calculations of the trigluon correlation function.  

The rest of the theory curves in Figure~\ref{pi0Theory} show predictions for the midrapidity $\pi^0$ TSSA calculated in the TMD picture.  These curves show the predicted asymmetry as generated by the Sivers TMD PDF for both quarks and gluons.  They are calculated using functions published in Ref~\cite{gluonSivers} which have been reevaluated at Feynman-$x$ ($x_F = 2 p_L/\sqrt{s}$) equal to zero,
which approximates the measured kinematics.  The red curve was calculated using the Generalized Parton Model (GPM) meaning that the first $k_T$ moment of the Sivers function is taken and the calculation does not include next-to-leading-order interactions with the proton fragment.  The color-gauge invariant generalized parton model (CGI-GPM) somewhat relaxes these assumptions  and includes both initial- and final-state interactions via the one-gluon exchange approximation.   
%This model has been shown to reproduce the predicted sign change for the quark Sivers function in SIDIS and Drell-Yan. 
The CGI-GPM curves plotted in Figure~\ref{pi0Theory} come from simultaneous fits to open heavy flavor~\cite{openHeavyFlavor} and midrapidity $\pi^0$~\cite{PPG135} TSSA measurements from PHENIX.  Scenario 1 maximizes the asymmetry, while scenario 2 minimizes it.  As shown in the zoomed in top panel of Figure~\ref{pi0Theory}, this $\pi^0$ TSSA measurement has the statistical precision to distinguish between the GPM and CGI-GPM models, preferring CGI-GPM Scenario 2.  

\section{Direct Photon TSSA}
Direct photons are photons that come \textit{directly} from the hard-scattering event.  Because they do not undergo hadronization they are only sensitive to initial-state effects  from proton structure.  At large transverse momentum they are produced by the 2-to-2 hard scattering subprocesses quark-gluon Compton scattering ($g + q \rightarrow \gamma + q$) and quark-antiquark annihilation ($\bar{q} + q \rightarrow \gamma + g$).  At midrapidity quark-gluon Compton scattering dominates~\cite{directPhotonProduction} because the proton is probed at a moderate $x$ where the gluon PDF dominates.  As a result, midrapidity direct photons are a uniquely clean probe of gluon structure in the proton.  

The vast number of photons present in an event are not direct photons, but instead come from decays and next-to-leading order fragmentation processes.  Many of these photons are eliminated by a tagging cut which removes that photons that have been matched into a pair with another photon in the same event which reconstructs either a $\pi^0 \rightarrow \gamma \gamma$ or $\eta \rightarrow \gamma \gamma$ decay.  An isolation cut further reduces the contribution of decay photons~\cite{PPG136} as well as next-to-leading order fragmentation photons to about 15\% for direct photons with $p_{T} > 5~\rm{GeV}/c$~\cite{directPhotonProduction}.  This cut adds up the energy of all of the surrounding EMCal clusters and the momentum of all of the surrounding tracks that are within a cone of 0.4 radians.  The photon only passes this isolation cut if it has ten times the energy of the surrounding cone.  The remaining background is dominated by decay photons where the second partner photon was missed because it was either out of acceptance or too low in energy.  This decay photon would not have been eliminated by the tagging cut so this background is estimated through single particle Monte Carlo and found to be about 50\% for the lowest $p_T$ bin and drop to about 16\% in the highest $p_T$ bin.  

\begin{figure}[h]
\centering
\includegraphics[width=0.7\textwidth]{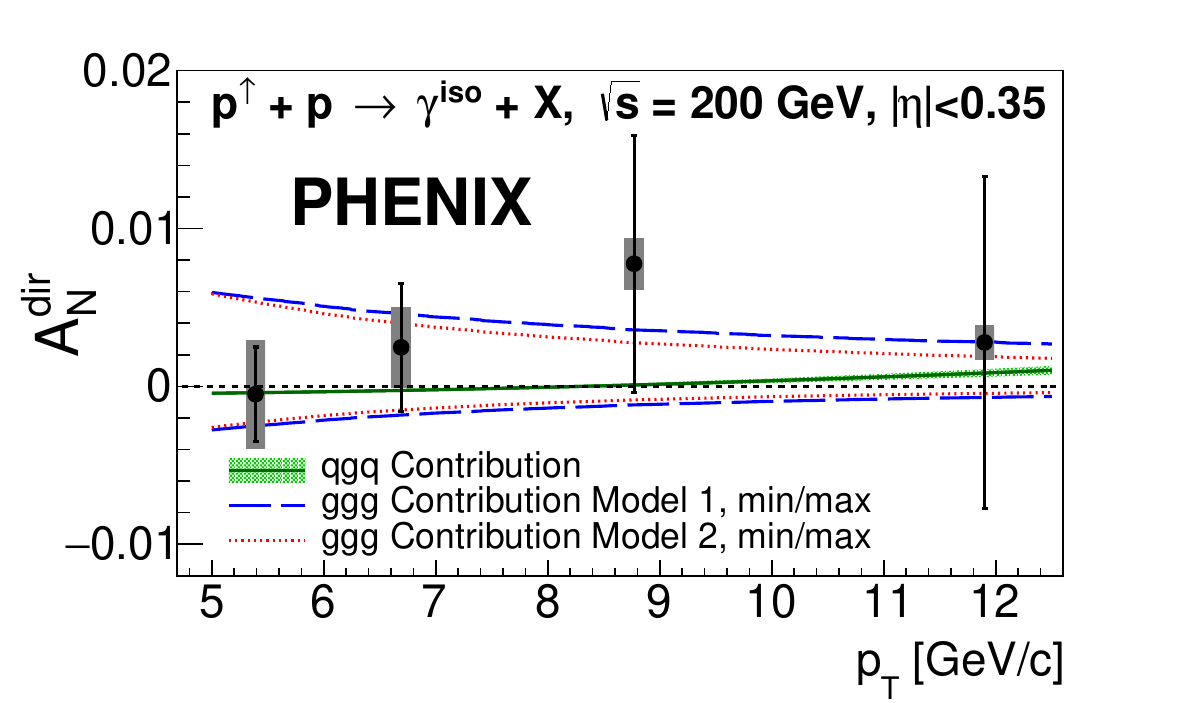}
\caption{The direct photon TSSA~\cite{PPG235} plotted with the contributions from the $qgq$~\cite{qgqDirectPhoton} and $ggg$~\cite{gggDirectPhoton} correlation function contributions.  See text for details. An additional scale uncertainty of 3.4\% due to the uncertainty in the polarization is not shown.}
\label{directPhoton}
\end{figure}

The final direct asymmetry is plotted in Figure~\ref{directPhoton} and is consistent with zero to within about 2\%~\cite{PPG235}.  The only previously published  direct photon TSSA result was measured at the E704 Fermilab experiment and was found to be consistent with zero to within about 20\% for $2.5 < p_T^\gamma < 3.1~\rm{GeV}/c$ for $\sqrt{s} = 19.4~\rm{GeV}$ ~\cite{oldDirectPhotonAsymmetry}.  This new result measured photons with $p_T^\gamma > 5~\rm{GeV}/c$ with total uncertainties a factor of 50 times smaller than the E704 measurement.  
The green curve in Figure~\ref{directPhoton} shows the contribution from $qgq$ correlation functions from both the polarized and unpolarized proton that were published in Ref.~\cite{qgqDirectPhoton} and recalculated for $|\eta|<0.35$.  The error bars shown correspond to propagated uncertainties from fits to data and do not include uncertainties from assuming functional forms.  The $ggg$ correlation function contributions use fits that were published in Ref.~\cite{gggDirectPhoton} and were reevaluated for $\eta = 0$.  Models 1 and 2 correspond to   different functional forms assumptions for the trigluon correlation function in  terms of the collinear leading-twist gluon PDF.  The trigluon correlation function is divided into symmetric and antisymmetric parts.  Setting these parts to have the same sign maximizes the direct photon asymmetry while setting them to have the opposite sign minimizes it.  Given the small predicted $qgq$ correlation function contribution, this direct photon asymmetry result will help constrain the trigluon correlation function.  

%In the list of references, cited papers \cite{1931_Bethe_ZP_71} should include authors, title, journal reference (journal name, volume number (in bold), start page) and most importantly a DOI link. For a preprint \cite{arXiv:1108.2700}, please include authors, title (please ensure proper capitalization) and arXiv link. If you use BiBTeX with our style file, the right format will be automatically implemented.

%All equations and references should be hyperlinked to ensure ease of navigation. This also holds for [sub]sections: readers should be able to easily jump to Section \ref{sec:another}.

%There is no strict length limitation, but the authors are strongly encouraged to keep contents to the strict minimum necessary for peers to reproduce the research described in the paper.

%Figures should only occupy the strictly necessary space, in any case individually fitting on a single page. Each figure item should be appropriately labeled and accompanied by a descriptive caption. {\bf SciPost does not accept creative or promotional figures or artist's impressions}; on the other hand, technical drawings and scientifically accurate representations are encouraged.

\section{Conclusion}
TSSAs are spin-momentum correlations that probe parton dynamics in the proton as well as in the process of hadronization.  They can be described by both the TMD and collinear twist-3 frameworks, were collinear twist-3 correlation functions only require a single hard-energy scale to be measured directly.  The midrapidity $\pi^0$ and $\eta$ TSSAs measurements were presented for $\sqrt{s} = 200~\rm{GeV}$.  Both results are consistent with zero and have a factor of 3 increase in precision compared to the previous PHENIX results.  Midrapidity $\pi^0$ and $\eta$ mesons are sensitive to both initial- and final-state effects for both quarks and gluons and the $\eta$ TSSA is in particular sensitive to strangeness in twist-3 functions.  Midrapidity isolated direct photons offer a clean probe of gluon initial-state effects.  The direct photon TSSA has been measured for the first time at RHIC and is also consistent with zero.  These asymmetry results will help constrain the trigluon correlation function in the transversely polarized proton as well as the gluons Sivers function, both of which are steps towards creating a more complete, three-dimensional picture of proton structure. %These measurements used PHENIX's final polarized proton-proton data set.  
%%%%%Fix last sentence

\section*{Acknowledgements}
Thank you to Daniel Pitonyak for providing the $qgq$ correlation function curves that appear in both Figures~\ref{pi0Theory} and ~\ref{directPhoton}.  Thank you to Umberto D’Alesio, Cristian Pisano, and Francesco Murgia for providing the GPM and CGI-GPM curves that are plotted in Figure~\ref{pi0Theory}.  Thank you to Shinsuke Yoshida for providing the $ggg$ correlation function curves as shown in Figure~\ref{directPhoton}.

%Acknowledgements should follow immediately after the conclusion.

% TODO: include funding information
\paragraph{Funding information}
The funding for this project was proved by the Department of Energy, grant number DE-SC0013393.

%Authors are required to provide funding information, including relevant agencies and grant numbers with linked author's initials. Correctly-provided data will be linked to funders listed in the \href{https://www.crossref.org/services/funder-registry/}{\sf Fundref registry}.

%\section{About references}
%Your references should start with the comma-separated author list (initials + last name), the publication title in italics, the journal reference with volume in bold, start page number, publication year in parenthesis, completed by the DOI link (linking must be implemented before publication). If using BiBTeX, please use the style files provided  on \url{https://scipost.org/submissions/author_guidelines}. If you are using our \LaTeX template, simply add
%\begin{verbatim}
%\bibliography{your_bibtex_file}
%\end{verbatim}
%at the end of your document. If you are not using our \LaTeX template, please still use our bibstyle as
%\begin{verbatim}
%\bibliographystyle{SciPost_bibstyle}
%\end{verbatim}
%in order to simplify the production of your paper.

% \bibliographystyle{SciPost_bibstyle} % Include this style file here only if you are not using our template
%\bibliography{SciPost_Example_BiBTeX_File.bib}

%\nolinenumbers

\end{document}